\documentclass[prd,twocolumn,showpacs,floatfix,amsmath,nofootinbib,amssymb,floatfix]{revtex4}
\usepackage{bbding}
\usepackage{graphicx,color,dcolumn,booktabs,bm}
\usepackage{longtable,lscape}
\usepackage{txfonts}
\usepackage{rotating}
\usepackage{overpic}
\usepackage{amssymb}
\usepackage{indentfirst}
\usepackage{feynmf}   
\usepackage{slashed}  
\usepackage{cases}
\usepackage{color}
\usepackage{multirow}
\usepackage{epstopdf}
\usepackage{longtable}
\usepackage{graphicx,color,dcolumn,booktabs,bm}
\usepackage[colorlinks,
            citecolor=blue,
            anchorcolor=red,
            menucolor=red,
            linkcolor=red,
            filecolor=red,
            runcolor=red,
            urlcolor=blue,
            frenchlinks=red]{hyperref}

\graphicspath{{Figures/}} %

\allowdisplaybreaks

\begin{document}

\title{Predicting possible molecular states of nucleons with $\Xi_c$, $\Xi_c^{*}$, and $\Xi_c^{\prime}$ }

\author{Jin-Yu Huo$^{1}$}
\author{Li-Cheng Sheng$^{1}$}
\author{Rui Chen$^{1}$\footnote{Corresponding author}}\email{chenrui@hunnu.edu.cn}
\author{Xiang Liu$^{2,3,4,5,6}$}
\email{xiangliu@lzu.edu.cn}

\affiliation{
$^1$Key Laboratory of Low-Dimensional Quantum Structures and Quantum Control of Ministry of Education, Department of Physics and Synergetic Innovation Center for Quantum Effects and Applications, Hunan Normal University, Changsha 410081, China\\
$^2$School of Physical Science and Technology, Lanzhou University, Lanzhou 730000, China\\
$^3$Lanzhou Center for Theoretical Physics, Key Laboratory of Theoretical Physics of Gansu Province, Lanzhou University, Lanzhou 730000, China\\
$^4$Key Laboratory of Quantum Theory and Applications of MoE, Lanzhou University, Lanzhou 730000, China\\
$^5$MoE Frontiers Science Center for Rare Isotopes, Lanzhou University, Lanzhou 730000, China\\
$^6$Research Center for Hadron and CSR Physics, Lanzhou University and Institute of Modern Physics of CAS, Lanzhou 730000, China}
\date{\today}

\begin{abstract}
In the framework of a one-boson-exchange model, we carry out a comprehensive investigation of the $\Xi_cN/\Lambda_c\Sigma/\Xi_c^{\prime}N/\Sigma_c\Lambda/\Xi_c^*N/\Sigma_c^*\Lambda/\Sigma_c\Sigma/\Sigma_c^*\Sigma$  interactions. We consider the $S$-$D$-wave mixing effects and the coupled-channel effects to derive the relevant effective potentials. Our results can predict several possible charm-strange deuteronlike $\Xi_c^{(',*)}N$ hexaquarks, the $\Xi_c^{\prime}N$ molecules with $I(I^P)=0(0^+)$, $0(1^+)$, $1(1^+)$, the $\Xi_c^*N$ molecules with $0(1^+)$, $0(2^+)$, $1(2^+)$, and the coupled $\Xi_cN/\Xi_c^{\prime}N/\Xi_c^*N/\Sigma_c\Sigma/\Sigma_c^*\Sigma$ molecule with $0(1^+)$. We expect the experiments to search for our predictions of the $\Xi_c^{(\prime,*)}N$ bound states.
\end{abstract}


\maketitle

\section{introduction}

As is well known, the extensive experimental data on nucleon-nucleon interactions have led to a profound understanding of the nuclear force, allowing a comprehensive exploration of the properties for nuclear matter \cite{Gomes:1957zz,Krein:2016fqh,Moszkowski:1960zz,Bethe:1971xm,Jackson:1983wfn,Brockmann:1990cn}. Extending this line of investigation to scenarios, where a light quark is replaced by a strange quark, paves the way for an in-depth study of hyperon-nucleon interactions ($YN$, with $Y$ representing $\Lambda$ or $\Sigma$). Such investigations may not only have the potential to predict the existence of $YN$ hypernuclei, but also provide valuable insights into the properties of strange baryons within nuclear matter (as detailed in reviews \cite{Hosaka:2016ypm,Krein:2017usp}).

The hypernuclei in the charmed sector might date back to 1977 \cite{Dover:1977jw}; the authors studied the $Y_cN$  $(Y_c=\Lambda_c, \Sigma_c)$ interactions by extending the meson-exchange model in an $SU(4)$ symmetry  and found that there should exist both two-body and many-body bound states of a charmed baryon and nucleons. Subsequently, many groups continued to study such $Y_cN$ hypernuclei by using various models, such as the Woods-Saxon form phenomenological potential well \cite{Cai:2003ce}, the quark-meson coupling model \cite{Tsushima:2003dd}, the chiral soliton models \cite{Kopeliovich:2007kd}, the chiral constituent quark model \cite{Garcilazo:2015qha,Epelbaum:2014sea}, the meson-exchange model in the heavy quark limit \cite{Liu:2011xc,Maeda:2015hxa,Maeda:2018xcl}, the HAL QCD method \cite{Miyamoto:2017tjs,Miyamoto:2017ynx}, and the chiral effective field theory \cite{Haidenbauer:2017dua,Meng:2019nzy}.

Recently, the E07 Collaboration at J-PARC observed a $\Xi^-$ absorption event decaying into twin single $\Lambda$ hypernuclei, $\Xi^-+{}^{14}N\to {}_{\Lambda}^{10}\text{Be}+{}_{\Lambda}^5\text{He}$, and deduced the binding energy of the $\Xi^-$ hyperon in the $\Xi^-$-${}^{14}N$ system $E=1.27\pm0.21$ MeV \cite{J-PARCE07:2020xbm}. With this finding, it is natural to study the formation of the charmed-strange hypernuclei, which is closely related to the properties of the interactions between the charmed-strange baryon and the nucleon.
Compared to the $\Xi N$ interactions, it is reasonable to expect that the attraction between $\Xi_c $ and $N$ might be slightly weaker due to the presence of fewer light quarks in the $\Xi_cN$ system. However, it is crucial to remember that the $\Xi_cN$ system is significantly heavier than the $\Xi N$ system. This increased mass results in a heavier reduced mass, which is favorable for the formation of a bound state due to the suppression of kinematic energy. It is therefore essential to carry out a quantitative calculation to explore the possible existence of a $\Xi_cN$ bound state. We have reason to believe that this study can enrich our knowledge of charmed hypernuclei.

Although charmed-strange hypernuclei are a focal topic of nuclear physics, they can be covered by hadron physics, since the system under discussion is a typical bound state problem, namely, the molecular states composed of charmed-strange baryon and nucleon.
Especially, in the last two decades, more and more candidates for exotic hadronic matter have been reported, including charmoniumlike $XYZ$ states, hidden-charm pentaquarks $P_{\psi}^N/P_{\psi}^{\Lambda}$, and doubly charmed tetraquark $T_{cc}$, inspiring extensive discussions on different types of exotic hadronic states (see reviews \cite{Chen:2016qju,Liu:2019zoy,Chen:2016spr,Guo:2017jvc,Chen:2022asf,Liu:2013waa,Hosaka:2016pey} for more details), with molecular states being the most popular among those involving hadronic configurations. Various methods and approaches have been developed to deal with the bound state problem (see \cite{Chen:2016qju,Guo:2017jvc,Chen:2022asf} for recent reviews).

In this work, we employ the one-boson-exchange (OBE) model \cite{Yukawa:1935xg,Machleidt:1987hj,Stoks:1994wp,Wiringa:1994wb} to derive the relevant effective potentials, a method often used to unravel the nature of these reported new hadronic states \cite{Chen:2015loa,Chen:2016ypj,Chen:2019asm,Chen:2019uvv,Chen:2021vhg,Chen:2020kco,Chen:2020yvq,Chen:2022dad,
Liu:2019tjn,He:2019ify,Yamaguchi:2019seo,Burns:2019iih}. This model takes into account the contributions of different exchanged mesons, namely the $\pi$, $\sigma$, and $\rho/\omega$ mesons, which are responsible for long-, intermediate-, and short-range interactions, respectively. In this context, the $\pi$ exchange interaction plays a crucial role in binding the well-known deuteron.

However, when it comes to $\Xi_cN$ interactions, they show significant differences compared to $NN$ interactions. First, the one-pion-exchange(OPE) interactions are strongly suppressed in the $\Xi_cN$ systems. This suppression results from the conservation of the light quark spin parity, which forbids the $\Xi_c-\Xi_c-\pi$ vertices. On the other hand, the coupled-channel effects play a significant role in $\Xi_cN$ interactions due to the proximity of the mass thresholds for the $S$-wave ground charm-strange baryons, namely $(\Xi_c, \Xi_c^{\prime}, \Xi_c^{\ast})$ \cite{ParticleDataGroup:2022pth}. For example, in Ref. \cite{Liu:2011xc}, the coupled-channel effects from the $\Sigma_cN$ and $\Sigma_c^*N$ channels are essential to generate the $\Lambda_cN$ bound state. As shown by PDG \cite{ParticleDataGroup:2022pth}, the spin parities for the $\Xi_c$, $\Xi_c^{\prime}$, and $\Xi_c^{\ast}$ are $J^P=1/2^+$, $1/2^+$, and $3/2^+$, respectively. The light quarks of the $\Xi_c$, $\Xi_c^{\prime(\ast)}$ are in flavor antisymmetry and symmetry, respectively. The corresponding spin parities of the light quarks $j_{qq}^P$ are $0^+$, and $1^+$, respectively. Their mass thresholds are very close, i.e., $M_{\Xi_c^{\prime}}-M_{\Xi_c}\simeq100$ and $M_{\Xi_c^*}-M_{\Xi_c}\simeq170$ MeV.

In our study, we take into account both the $S$-$D$-wave mixing effects and the coupled-channel effects when deducing the $\Xi_c^{(\prime,*)}N$ interactions. In addition to studying the coupled $\Xi_cN/\Xi_c^{\prime}N/\Xi_c^*N$ interactions, we further discuss the coupling from the $\Lambda_c\Sigma$, $\Sigma_c^{(*)}\Lambda$, and $\Sigma_c^{(*)}\Sigma$ channels as their masses are also close to the $\Xi_c^{(\prime,*)}N$ thresholds. Using these derived OBE effective potentials, we proceed to search for bound state solutions by solving the coupled-channel Schr\"{o}dinger equations. Ultimately, our investigation aims to answer the question of whether charm-strange deuteronlike $\Xi_c^{(\prime,*)}N$ hexaquarks could potentially exist. Indeed, the results of this investigation may not only provide invaluable guidance at the discovery of potential charm-strange hypernuclei but also expand our knowledge of the rich and intricate world of hypernuclei, shedding some light on the behavior of charm and strangeness within atomic nuclei to some extent.

This paper is organized as follows. After this Introduction, in Sec.~\ref{sec2} we present the OBE effective potentials for the discussed systems. In Sec.~\ref{sec3}, we present the corresponding numerical results for the $\Xi_c^{(\prime,*)}N$ systems. The paper ends with a summary in Sec. \ref{sec4}.

\section{The OBE effective potentials}\label{sec2}

Before deriving the OBE effective potentials for the $\Xi_c^{(',*)}N$ systems, we first construct the wave functions. These wave functions can be expressed as the direct product of the spin-orbit wave function $|{}^{2S+1}L_J\rangle$, the flavor wave function $|I,I_3\rangle$, and the spatial wave function $\psi(r)$. The isospin for the $\Xi_c^{(',*)}N$ systems can be either 0 or 1. The flavor wave functions for the isoscalar $\Xi_c^{(',*)}N$ systems are defined as $|0,0\rangle=(|\Xi_c^{(',*)+}n\rangle-|\Xi_c^{(',*)0}p\rangle)/\sqrt{2}$. For the isovector $\Xi_c^{(',*)}N$ systems, the flavor wave functions are $|1,1\rangle=|\Xi_c^{(',*)+}p\rangle$, $|1,0\rangle=(|\Xi_c^{(',*)+}n\rangle+|\Xi_c^{(',*)0}p\rangle)/\sqrt{2}$, and  $|1,-1\rangle=|\Xi_c^{(',*)0}n\rangle$.

After taking into account the $S$-$D$-wave mixing effects, we construct the spin-orbit wave functions for the $\Xi_c^{(',*)}N$ systems as follows:
\begin{eqnarray}
\Xi_{c}^{(')}N:  &J^P=0^+&  \left|{}^1S_0\right\rangle,\label{spin1}\\
                           &J^P=1^+&  \left|{}^3S_1\right\rangle,\quad   \left|{}^3D_1\right\rangle,\\
\Xi_{c}^*N:  &J^P=0^+&  \left|{}^5D_0\right\rangle,\\
&J^P=1^+&  \left|{}^3S_1\right\rangle,\quad   \left|{}^3D_1\right\rangle,\quad
                                      \left|{}^5D_1\right\rangle,\\
                           &J^P=2^+&  \left|{}^5S_2\right\rangle,\quad   \left|{}^3D_2\right\rangle,\quad
                                      \left|{}^5D_2\right\rangle.\label{spin4}
\end{eqnarray}
The general expressions for the spin-orbit wave functions for the $\Xi_c^{(',*)}N$ systems are as follows:
\begin{eqnarray*}
\Xi_{c}^{(')}N \left|{}^{2S+1}L_{J}\right\rangle &=& \sum_{m,n,m_L}C_{\frac{1}{2},m;\frac{1}{2},n}^{S,ms}C_{S,m_S;L,m_L}^{J,m_J}
       \chi_{\frac{1}{2},m}\chi_{\frac{1}{2},n}Y_{L,m_L},\\
\Xi_{c}^*N \left|{}^{2S+1}L_{J}\right\rangle &=&
\sum_{m,n,m_L}C_{\frac{1}{2},m;\frac{3}{2},n}^{S,m_S}C_{S,m_S;L,m_L}^{J,m_J}
       \chi_{\frac{1}{2},m}\Phi_{\frac{3}{2},n}Y_{L,m_L}.
\end{eqnarray*}
Here, $C_{\frac{1}{2},m;\frac{1}{2},n}^{S,ms}$, $C_{\frac{1}{2},m;\frac{3}{2},n}^{S,m_S}$, and $C_{S,m_S;L,m_L}^{J,m_J}$ are the Clebsch-Gordan coefficients. $\chi_{\frac{1}{2},m}$ and $Y_{L,m_L}$ denote the spin wave function for the baryon with $J=1/2$ and the spherical harmonics function, respectively. $\Phi_{\frac{3}{2},n}=\sum_{n_1,n_2}\langle\frac{1}{2},n_1;1,n_2|\frac{3}{2},n\rangle
\chi_{\frac{1}{2},n_1}\epsilon^{n_2}$ is the spin wave function for the baryon with $J=3/2$, where $\epsilon$ stands for the polarization vector, $\epsilon_{\pm1}= \frac{1}{\sqrt{2}}\left(0,\pm1,i,0\right)$ and $\epsilon_{0} =\left(0,0,0,-1\right)$.

The general procedures for deducing the OBE effective potentials can be divided into three steps. First, one can write down the $t$-channel scattering amplitudes $\mathcal{M}(h_1h_2\to h_3h_4)$ based on the corresponding effective Lagrangians. Second, one can derive the OBE effective potentials in the momentum space by using a Breit approximation, i.e.,
\begin{eqnarray}
\mathcal{V}_{E}^{h_1h_2\to h_3h_4}(\vec{q}) &=&
          -\frac{\mathcal{M}(h_1h_2\to h_3h_4)}
          {\sqrt{\prod_i2p_i^0\prod_f2p_f^0}}.
\end{eqnarray}
Here, $p_i^0$ and $p_f^0$ are the energies of the initial states and final states, respectively. Finally, the effective potentials in the coordinate space $\mathcal{V}(r)$ can be obtained by performing a Fourier transformation,
\begin{eqnarray}
\mathcal{V}_{E}^{h_1h_2\to h_3h_4}({r}) =
          \int\frac{d^3\vec{q}}{(2\pi)^3}e^{i\vec{q}\cdot\vec{r}}
          \mathcal{V}_{E}^{h_1h_2\to h_3h_4}(\vec{q})\mathcal{F}^2(q^2,m_E^2).
\end{eqnarray}
To reflect the finite size effect of the hadrons involved in these interactions, we have introduced a monopole form factor $\mathcal{F}(q^2,m_E^2)=(\Lambda^2-m_E^2)/(\Lambda^2-q^2)$ in each interaction vertex. $m_E$ and $q$ stand for the mass and four-momentum of the exchanged mesons, respectively. The cutoff $\Lambda$ is a free parameter, which is related to the typical hadronic scale or the intrinsic size of hadrons.

In the present study, we assume that the intrinsic size of the charm-strange baryon is similar to that of the nucleon. In the following, we search for the bound state solutions by varying the cutoff $\Lambda$ from 0.80 to 2.00 GeV, using the similar cutoff $\Lambda$ of the deuteron, $\Lambda\sim$ 1.00 GeV \cite{Tornqvist:1993ng,Tornqvist:1993vu} is set as a reasonable input.

The effective Lagrangians describing the interactions between the charmed baryons and the light meson interactions can be constructed on the basis of the heavy quark symmetry, the chiral symmetry and the hidden local symmetry \cite{Liu:2011xc}, which are given by
\begin{eqnarray}
\mathcal{L}_{\mathcal{B}_{\bar{3}}} &=& l_B\langle\bar{\mathcal{B}}_{\bar{3}}\sigma\mathcal{B}_{\bar{3}}\rangle
          +i\beta_B\langle\bar{\mathcal{B}}_{\bar{3}}v^{\mu}(\mathcal{V}_{\mu}-\rho_{\mu})\mathcal{B}_{\bar{3}}\rangle,\label{lag1}\\
\mathcal{L}_{\mathcal{B}_{6}} &=&  l_S\langle\bar{\mathcal{S}}_{\mu}\sigma\mathcal{S}^{\mu}\rangle
         -\frac{3}{2}g_1\varepsilon^{\mu\nu\lambda\kappa}v_{\kappa}
         \langle\bar{\mathcal{S}}_{\mu}\mathcal{A}_{\nu}\mathcal{S}_{\lambda}\rangle\nonumber\\
         &&+i\beta_{S}\langle\bar{\mathcal{S}}_{\mu}v_{\alpha}\left(\mathcal{V}^{\alpha}-\rho^{\alpha}\right) \mathcal{S}^{\mu}\rangle
         +\lambda_S\langle\bar{\mathcal{S}}_{\mu}F^{\mu\nu}(\rho)\mathcal{S}_{\nu}\rangle,\label{lag2}\nonumber\\\\
\mathcal{L}_{\mathcal{B}_{\bar{3}}\mathcal{B}_6} &=& ig_4\langle\bar{\mathcal{S}^{\mu}}\mathcal{A}_{\mu}\mathcal{B}_{\bar{3}}\rangle+i\lambda_I\varepsilon^{\mu\nu\lambda\kappa}v_{\mu}\langle \bar{\mathcal{S}}_{\nu}F_{\lambda\kappa}\mathcal{B}_{\bar{3}}\rangle+H.c.\label{lag03}
\end{eqnarray}
Here, $v=(1,\textbf{0})$, $F^{\mu\nu}(\rho)=\partial^{\mu}\rho^{\nu}-\partial^{\nu}\rho^{\mu}
+\left[\rho^{\mu},\rho^{\nu}\right]$ with $\rho^{\mu}=ig_V{V}^{\mu}/\sqrt{2}$. $\mathcal{A}_{\mu}=\frac{1}{2}(\xi^{\dag}\partial_{\mu}\xi-\xi\partial_{\mu}\xi^{\dag})$ and $\mathcal{V}_{\mu}=\frac{1}{2}(\xi^{\dag}\partial_{\mu}\xi+\xi\partial_{\mu}\xi^{\dag})$ are the axial current and vector current, respectively. $\xi=\text{exp}(i{P}/f_{\pi})$ and $f_{\pi}=132$ MeV. The superfield $\mathcal{S}$ is a combination of heavy baryon fields $\mathcal{B}_6$ with $J^P=1/2^+$ and $\mathcal{B}^*_6$ with $J^P=3/2^+$, $\mathcal{S}_{\mu} =-\sqrt{\frac{1}{3}}(\gamma_{\mu}+v_{\mu})\gamma^5\mathcal{B}_6+\mathcal{B}_{6\mu}^*$. The matrices $\mathcal{B}_{\bar{3}}$, $\mathcal{B}_6$, ${P}$, and ${V}$ are written as
\begin{eqnarray*}
\mathcal{B}_{\bar{3}} &=& \left(\begin{array}{ccc}
        0    &\Lambda_c^+      &\Xi_c^+\\
        -\Lambda_c^+       &0      &\Xi_c^0\\
        -\Xi_c^+      &-\Xi_c^0     &0
\end{array}\right),\\
\mathcal{B}_6^{(*)} &=& \left(\begin{array}{ccc}
         \Sigma_c^{{(*)}++}                  &\frac{\Sigma_c^{{(*)}+}}{\sqrt{2}}     &\frac{\Xi_c^{(',*)+}}{\sqrt{2}}\\
         \frac{\Sigma_c^{{(*)}+}}{\sqrt{2}}      &\Sigma_c^{{(*)}0}    &\frac{\Xi_c^{(',*)0}}{\sqrt{2}}\\
         \frac{\Xi_c^{(',*)+}}{\sqrt{2}}    &\frac{\Xi_c^{(',*)0}}{\sqrt{2}}      &\Omega_c^{(*)0}
\end{array}\right),\\
P &=& \left(\begin{array}{ccc}
\frac{\pi^0}{\sqrt{2}}+\frac{\eta}{\sqrt{6}} &\pi^+ &K^+ \nonumber\\
\pi^- &-\frac{\pi^0}{\sqrt{2}}+\frac{\eta}{\sqrt{6}} &K^0 \nonumber\\
K^- &\bar{K}^0 &-\sqrt{\frac{2}{3}}\eta
\end{array}\right),\nonumber\\
V &=& \left(\begin{array}{ccc}
\frac{\rho^0}{\sqrt{2}}+\frac{\omega}{\sqrt{2}} &\rho^+ &K^{*+} \nonumber\\
\rho^- &-\frac{\rho^0}{\sqrt{2}}+\frac{\omega}{\sqrt{2}} &K^{*0} \nonumber\\
K^{*-} &\bar{K}^{*0} &\phi
\end{array}\right),
\end{eqnarray*}
respectively.

By expanding Eqs. (\ref{lag1})$-$(\ref{lag03}), we can obtain the concrete effective Lagrangians, i.e.,
\begin{widetext}
\begin{eqnarray}
\mathcal{L}_{\mathcal{B}_{\bar{3}}\mathcal{B}_{\bar{3}}\sigma} &=& l_B\langle \bar{\mathcal{B}}_{\bar{3}}\sigma\mathcal{B}_{\bar{3}}\rangle,\\\nonumber\\
\mathcal{L}_{\mathcal{B}_{6}^{(*)}\mathcal{B}_{6}^{(*)}\sigma} &=& -l_S\langle\bar{\mathcal{B}}_6\sigma\mathcal{B}_6\rangle
+l_S\langle\bar{\mathcal{B}}_{6\mu}^{*}\sigma\mathcal{B}_6^{*\mu}\rangle,\\\nonumber\\
\mathcal{L}_{\mathcal{B}_{\bar{3}}\mathcal{B}_{\bar{3}}{V}} &=& \frac{1}{\sqrt{2}}\beta_Bg_V\langle\bar{\mathcal{B}}_{\bar{3}}
v\cdot{V}\mathcal{B}_{\bar{3}}\rangle,\\\nonumber\\
\mathcal{L}_{\mathcal{B}_6^{(*)}\mathcal{B}_6^{(*)}{P}} &=&
        i\frac{g_1}{2f_{\pi}}\varepsilon^{\mu\nu\lambda\kappa}v_{\kappa}
        \langle\bar{\mathcal{B}}_6
        \gamma_{\mu}\gamma_{\lambda}\partial_{\nu}{P}\mathcal{B}_6\rangle\nonumber-i\frac{3g_1}{2f_{\pi}}\varepsilon^{\mu\nu\lambda\kappa}v_{\kappa}\langle
\bar{\mathcal{B}}_{6\mu}^{*}\partial_{\nu} {P}\mathcal{B}_{6\lambda}^*\rangle\nonumber+i\frac{\sqrt{3}}{2}\frac{g_1}{f_{\pi}}v_{\kappa}\varepsilon^{\mu\nu\lambda\kappa}
      \langle\bar{\mathcal{B}}_{6\mu}^*\partial_{\nu}{P}{\gamma_{\lambda}\gamma^5}
      \mathcal{B}_6\rangle+H.c.,\\\\
\mathcal{L}_{\mathcal{B}_6^{(*)}\mathcal{B}_6^{(*)} {V}} &=& -\frac{\beta_Sg_V}{\sqrt{2}}\langle\bar{\mathcal{B}}_6v\cdot{V}\mathcal{B}_6\rangle\nonumber-i\frac{\lambda g_V}{3\sqrt{2}}\langle\bar{\mathcal{B}}_6\gamma_{\mu}\gamma_{\nu}
    \left(\partial^{\mu} {V}^{\nu}-\partial^{\nu} {V}^{\mu}\right)
    \mathcal{B}_6\rangle\nonumber-\frac{\beta_Sg_V}{\sqrt{6}}\langle\bar{\mathcal{B}}_{6\mu}^*v\cdot {V}\left(\gamma^{\mu}+v^{\mu}\right)\gamma^5\mathcal{B}_6\rangle\nonumber\\
    &&-i\frac{\lambda_Sg_V}{\sqrt{6}}\langle\bar{\mathcal{B}}_{6\mu}^*
    \left(\partial^{\mu} {V}^{\nu}-\partial^{\nu} {V}^{\mu}\right)
    \left(\gamma_{\nu}+v_{\nu}\right)\gamma^5\mathcal{B}_6\rangle\nonumber+\frac{\beta_Sg_V}{\sqrt{2}}\langle\bar{\mathcal{B}}_{6\mu}^*v\cdot {V}\mathcal{B}_6^{*\mu}\rangle\nonumber\\
    &&+i\frac{\lambda_Sg_V}{\sqrt{2}}\langle\bar{\mathcal{B}}_{6\mu}^*
    \left(\partial^{\mu} {V}^{\nu}-\partial^{\nu} {V}^{\mu}\right)
    \mathcal{B}_{6\nu}^*\rangle+H.c.,\\ \nonumber\\
\mathcal{L}_{\mathcal{B}_{\bar{3}}\mathcal{B}_6^{(*)}{V}} &=&
       -\frac{\lambda_Ig_V}{\sqrt{6}}\varepsilon^{\mu\nu\lambda\kappa}v_{\mu}\langle \bar{\mathcal{B}}_6\gamma^5\gamma_{\nu}
        \left(\partial_{\lambda} {V}_{\kappa}-\partial_{\kappa} {V}_{\lambda}\right)\mathcal{B}_{\bar{3}}\rangle\nonumber-\frac{\lambda_Ig_V}{\sqrt{2}}\varepsilon^{\mu\nu\lambda\kappa}v_{\mu}\langle \bar{\mathcal{B}}_{6\nu}^*
          \left(\partial_{\lambda}{V}_{\kappa}-\partial_{\kappa}{V}_{\lambda}\right)
          \mathcal{B}_{\bar{3}}\rangle+H.c.,\\ \\
\mathcal{L}_{\mathcal{B}_{\bar{3}}\mathcal{B}_6^{(*)} {P}} &=& -\sqrt{\frac{1}{3}}\frac{g_4}{f_{\pi}}\langle\bar{\mathcal{B}}_6\gamma^5
\left(\gamma^{\mu}+v^{\mu}\right)\partial_{\mu}{P}\mathcal{B}_{\bar{3}}\rangle\nonumber-\frac{g_4}{f_{\pi}}\langle\bar{\mathcal{B}}_{6\mu}^*\partial^{\mu} {P}\mathcal{B}_{\bar{3}}\rangle+H.c.\\
\end{eqnarray}
\end{widetext}

For the coupling constants relevant to the charmed baryon interactions, they can be related to the nucleon-nucleon interactions by using the quark model~\cite{Liu:2011xc}, where $l_S=-2l_B=-\frac{2}{3}g_{\sigma NN}$, $g_1=-\frac{2\sqrt{2}}{3}g_4=-\frac{4\sqrt{2}f_{\pi}g_{\pi NN}}{5 m_\pi}$, $\beta_{S}^\rho g_V=-2\beta_{B}^\rho g_V=-4g_{\rho NN}$, $\lambda_{S}^\rho g_V=-\sqrt{8}\lambda_{I}^\rho g_V=-\frac{6(g_{\rho NN}+f_{\rho NN})}{5M_N}, $ $\beta_{S}^\omega g_V=-2\beta_{B}^\omega g_V=-\frac{4}{3}g_{NN\omega}$, and $\lambda_{S}^\omega g_V=-\sqrt{8}\lambda_{I}^\omega g_V=-2\frac{(g_{\omega NN}+f_{\omega NN})}{M_N}$.

The effective Lagrangians for the light baryons are
\begin{eqnarray}
\mathcal{L}_{B B \sigma} &  = & g_{B B \sigma} \bar{B} \sigma B, \\
\mathcal{L}_{B B P} & = & \frac{g_{B B P}}{m_{P}} \bar{B} \gamma^{5} \gamma^{\mu} \partial_{\mu} P B, \\
\mathcal{L}_{B B V} & = & g_{B B V} \bar{B}\gamma^{\mu}V_{\mu}B-\frac{f_{B B V}}{2 m_{B}}\bar{B} \sigma^{\mu v} \partial_{v} V_{\mu} B.
\end{eqnarray}
Here, $B$ represents light baryons in SU(3) octet. With the above preparations, we can deduce the OBE effective potentials for the processes $\mathcal{B}(1)B(2)\to\mathcal{B}(3)B(4)$ as follows:
\begin{widetext}
\begin{eqnarray}
\mathcal{V}_{\mathcal{B}_{\bar{3}}B\to\mathcal{B}_{\bar{3}}B}^{I} &=&
  -\mathcal{G}_\sigma^I g_{\sigma BB}l_BY(\Lambda,m_{\sigma},r)+\mathcal{G}_V^I\frac{g_{BBV}\beta_Bg_V}{\sqrt{2}}
  Y(\Lambda,m_{V},r)+\mathcal{G}_V^I\frac{f_{BBV}\beta_Bg_V}{8\sqrt{2}m_B}(\frac{1}{m_1}+\frac{1}{m_3})\nabla^2
     Y(\Lambda,m_{V},r),\label{pot1}\\ \nonumber\\
\mathcal{V}^I_{\mathcal{B}_{6}B\to\mathcal{B}_{6}B} &=&
  \mathcal{G}_\sigma^I g_{\sigma BB}l_SY(\Lambda,m_{\sigma},r)-\mathcal{G}_P^I\frac{g_{BBP}g_1}{3f_{\pi}m_P}\mathcal{F}(r,\bm{\sigma}_1,\bm{\sigma}_2)
 Y(\Lambda,m_{P},r)\nonumber\\
 &&-\mathcal{G}_V^I\left(\frac{\beta_Sg_Vf_{BBV}}{8\sqrt{2}m_B}(\frac{1}{m_1}+\frac{1}{m_3})+\frac{\lambda_Sg_Vg_{BBV}}{6\sqrt{2} }(\frac{1}{m_2}+\frac{1}{m_4})\right)\nabla^2Y(\Lambda,m_{V},r)\nonumber\\
  &&-\mathcal{G}_V^I\frac{\beta_Sg_Vg_{BBV}}{\sqrt{2}}Y(\Lambda,m_{V},r)
  -\mathcal{G}_V^I\frac{\lambda_Sg_V}{9\sqrt{2} }(\frac{g_{BBV}}{2m_1}+\frac{g_{BBV}}{2m_3}+\frac{f_{BBV}}{m_B})\mathcal{F}^{\prime}(r,\bm{\sigma}_1,\bm{\sigma}_2)Y(\Lambda,m_{V},r),\label{pot2}\\ \nonumber\\
\mathcal{V}^I_{\mathcal{B}_{6}^*B\to\mathcal{B}_{6}^*B} &=&
  \mathcal{G}_\sigma^I g_{\sigma BB}l_S\sum_{a,b}^{m,n}C_{1/2,a;1,b}^{3/2,a+b}
  C_{1/2,m;1,n}^{3/2,m+n}\chi_{4a}^{\dag}\chi_{2m}
     \bm{\epsilon}_2^n\cdot\bm{\epsilon}_4^{b\dag}Y(\Lambda,m_{\sigma},r)\nonumber\\
     &&-\mathcal{G}_P^I\frac{g_{BBP}g_1}{2f_{\pi}m_P}\sum_{a,b}^{m,n}C_{{1}/{2},a;1,b}^{{3}/{2},a+b}
  C_{1/2,m;1,n}^{3/2,m+n}\chi_{4a}^{\dag}
  \mathcal{F}(r,\bm{\sigma}_1,i\bm{\epsilon}_2^n\times\bm{\epsilon}_4^{b\dag})\chi_{2m}
  Y(\Lambda,m_{P},r)\nonumber\\
  &&-\mathcal{G}_V^I\frac{\beta_Sg_Vg_{BBV}}{\sqrt{2}}\sum_{a,b}^{m,n}C_{1/2,a;1,b}^{3/2,a+b}
  C_{1/2,m;1,n}^{3/2,m+n}\chi_{4a}^{\dag}\chi_{2m}
     \bm{\epsilon}_2^n\cdot\bm{\epsilon}_4^{b\dag}Y(\Lambda,m_{V},r)\nonumber\\
  &&-\mathcal{G}_V^I\frac{\beta_Sg_Vf_{BBV}}{8\sqrt{2}m_B}(\frac{1}{m_1}+\frac{1}{m_3})\sum_{a,b}^{m,n}C_{1/2,a;1,b}^{3/2,a+b}
  C_{1/2,m;1,n}^{3/2,m+n}\chi_{4a}^{\dag}\chi_{2m}
     \bm{\epsilon}_2^n\cdot\bm{\epsilon}_4^{b\dag}\nabla^2
  Y(\Lambda,m_{V},r)\nonumber\\
  &&-\mathcal{G}_V^I\frac{\lambda_s g_V}{6\sqrt{2}}(\frac{g_{BBV}}{2m_1}+\frac{g_{BBV}}{2m_3}+\frac{f_{BBV}}{m_B})\sum_{a,b}^{m,n}C_{{1}/{2},a;1,b}^{{3}/{2},a+b}
  C_{1/2,m;1,n}^{3/2,m+n}\chi_{4a}^{\dag}
  \mathcal{F}^{\prime}(r,\bm{\sigma}_1,i\bm{\epsilon}_2^n\times\bm{\epsilon}_4^{b\dag})
  \chi_{2m}
  Y(\Lambda,m_{V},r),\label{pot3}\\ \nonumber\\
\mathcal{V}^I_{\mathcal{B}_{\bar{3}}B\to\mathcal{B}_{6}B} &=&
  -\mathcal{G}_P^I\frac{1}{3\sqrt{3}}\frac{g_{BBP}g_4}{f_{\pi}m_P}\mathcal{F}(r,\bm{\sigma}_1,\bm{\sigma}_2)
  Y(\Lambda_0,m_{P0},r)-\mathcal{G}_V^I\frac{\lambda_Ig_V}{3\sqrt{6}}(\frac{g_{BBV}}{2m_1}+\frac{g_{BBV}}{2m_3}+\frac{f_{BBV}}{m_B})\mathcal{F}^{\prime}(r,\bm{\sigma}_1,\bm{\sigma}_2)
  Y(\Lambda_0,m_{V0},r),\label{pot4}\\ \nonumber\\
\mathcal{V}^I_{\mathcal{B}_{\bar{3}}B\to\mathcal{B}_{6}^*B} &=&
   \mathcal{G}_P^I\frac{g_{BBP}g_4}{3f_{\pi}m_P}\sum_{m,n}C_{1/2,m;1,n}^{3/2,m+n}\chi_{4m}^{\dag}
\mathcal{F}(r,\bm{\sigma}_1,\bm{\epsilon}_4^{n\dag})
  Y(\Lambda_0,m_{P0},r)\nonumber\\
  &&+\mathcal{G}_{V}^I\frac{\lambda_Ig_V}{3\sqrt{2}}(\frac{g_{BBV}}{2m_1}+\frac{g_{BBV}}{2m_3}+\frac{f_{BBV}}{m_B})\sum_{m,n}C_{1/2,m;1,n}^{3/2,m+n}\chi_{4m}^{\dag}
\mathcal{F}^{\prime}(r,\bm{\sigma}_1,\bm{\epsilon}_4^{n\dag})
  Y(\Lambda_0,m_{V0},r),\label{pot5}\\ \nonumber\\
\mathcal{V}^I_{\mathcal{B}_{6}B\to\mathcal{B}_{6}^*B} &=& -\mathcal{G}_\sigma^I\frac{l_Sg_{\sigma BB}}{\sqrt{3}}\sum_{m,n}C_{1/2,m;1,n}^{3/2,m+n}\chi_{4m}^{\dag}
   (\bm{\sigma}_2\cdot\bm{\epsilon}_4^{n\dag})Y(\Lambda_0,m_{\sigma0},r)
   +\mathcal{G}_V^I\frac{\beta_Sg_Vg_{BBV}}{\sqrt{6}}\sum_{m,n}C_{1/2,m;1,n}^{3/2,m+n}\chi_{4m}^{\dag}
  (\bm{\sigma}_2\cdot\bm{\epsilon}_4^{n\dag})Y(\Lambda_0,m_{V0},r)\nonumber\\
  &&+\mathcal{G}_P^I\frac{g_{BBP}g_1}{2\sqrt{3}m_Pf_{\pi}}\sum_{m,n}C_{1/2,m;1,n}^{3/2,m+n}\chi_{4m}^{\dag}
\mathcal{F}(r,\bm{\sigma}_1,i\bm{\sigma}_2\times\bm{\epsilon}_4^{n\dag})
  Y(\Lambda_0,m_{P0},r)\nonumber\\
  &&+\mathcal{G}_V^I\frac{\lambda_Sg_V}{6\sqrt{6}}(\frac{g_{BBV}}{2m_1}+\frac{g_{BBV}}{2m_3}+\frac{f_{BBV}}{m_B})\sum_{m,n}C_{1/2,m;1,n}^{3/2,m+n}\chi_{4m}^{\dag}
\mathcal{F}^{\prime}(r,\bm{\sigma}_1,i\bm{\sigma}_2\times\bm{\epsilon}_4^{n\dag})
  Y(\Lambda_0,m_{V0},r).\label{pot6}\nonumber\\
\end{eqnarray}
\end{widetext}
In the above effective potentials, we define several useful functions, i.e.,
\begin{eqnarray}
S(\hat{r},\bm{a},\bm{b}) &=& 3(\hat{r}\cdot\bm{a})(\hat{r}\cdot\bm{b})-\bm{a}\cdot\bm{b},\\ \nonumber\\
\mathcal{F}(r,\bm{a},\bm{b}) &=& \bm{a}\cdot\bm{b}\nabla^2
     +S(\hat{r},\bm{a},\bm{b})
     r\frac{\partial}{\partial r}\frac{1}{r}\frac{\partial}{\partial r},\\ \nonumber\\
\mathcal{F}^{\prime}(r,\bm{a},\bm{b}) &=&
     2\bm{a}\cdot\bm{b}\nabla^2
     -S(\hat{r},\bm{a},\bm{b})
     r\frac{\partial}{\partial r}\frac{1}{r}\frac{\partial}{\partial r},\\ \nonumber\\
Y(\Lambda, m, r) &=& \left\{\begin{array}{l}
\frac{e^{-mr}-e^{-\Lambda r}}{4\pi r}-\frac{\Lambda^2-m^2}{8\pi \Lambda}e^{-\Lambda r}, \quad(\text{for}~m_E^{2}>q_i^2)\\ \nonumber\\
\frac{1}{4\pi r}\left(-e^{-\Lambda r}-\frac{(\Lambda^2+m^2)r}{2\Lambda}e^{-\Lambda r}+\text{cos}(mr)\right).\end{array}\right.\\
\end{eqnarray}
The variables in Eqs. (\ref{pot1})-(\ref{pot6}) are
$\Lambda_0^2=\Lambda^2-q_0^2$, $m_{E0}^2=|m_{E}^2-q_0^2|$. In Table \ref{num:parameter}, we collect the isospin factor for different exchanged mesons, the time component of the transferred momentum $q^0$, the coupling constants, and the masses for the involved hadrons adopted in our calculations. The $\bm{a}\cdot\bm{b}$ and $S(\hat{r},\bm{a},\bm{b})$ correspond to the spin-spin interactions and the tensor force operators, respectively. Sandwiching these operators between the corresponding spin-orbit wave functions, as shown in Eqs. (\ref{spin1})-(\ref{spin4}), one can obtain the corresponding matrix elements, which are summarized in Table \ref{operator}.

\renewcommand{\arraystretch}{2.0}
\begin{table*}[!hbtp]
\renewcommand\tabcolsep{0.17cm}
\renewcommand{\arraystretch}{1.7}
\caption{The isospin factor for different exchanged mesons and the time component of the transferred momentum $q^0$ adopted in our calculations. The coupling constants are given in Refs. \cite{Machleidt:2000ge,Machleidt:1987hj,Cao:2010km}. The masses for the involved hadrons are taken from the PDG \cite{ParticleDataGroup:2022pth}.} \label{num:parameter}
\begin{tabular}{c|cccccc|cc|cc}
\toprule[1pt]
\toprule[1pt]
Process                                       &$ \sigma$  &$\eta$  &$\omega$                             &$\pi,\rho$                                  &$K,K^*$       &$q^0$   &\multicolumn{2}{c|}{Coupling constants}  &Mass (MeV) \\
\midrule[0.65pt]
$\Xi_cN\to\Xi_cN$                            &$2$  &$...$  &$\frac{1}{\sqrt{2}}$                          &$\frac{1}{\sqrt{2}}(\frac{-3}{\sqrt{2}})$   &$...$           &0  &$\frac{g^2_{\sigma NN}}{4\pi}=5.69$  &$g_{NN\eta}=0.33$   &$m_\sigma=600.00$\\

$\Xi_cN\to\Lambda_c\Sigma$                   &$...$   &$...$    &$...$                                        &$...$                                         &$-\sqrt{2}$
      &$\frac{m_{\Xi_c}^2-m_N^2+m_\Sigma^2-m_{\Lambda_c}^2}{2m_{\Lambda_c}+2m_\Sigma}$   &$\frac{g^2_{\pi NN}}{4\pi}=0.07$   &$g_{\Lambda\Lambda\omega}=7.98$  &$m_\pi=137.27$\\\

$\Xi_cN\to\Xi_c^{(\prime,*)}N$               &$...$  &$ \frac{\sqrt{3}}{2}$  &$\frac{1}{2}$               &$\frac{1}{2}(\frac{-3}{2})$      &$...$
      &\multirow{1}{*}{$\frac{m_{\Xi_c}^2-m_{\Xi_c^{\prime,*}}^2}{2m_{\Xi_c^{\prime,*}}+2m_N}$}   &$\frac{g^2_{\rho NN}}{4\pi}=0.81$ &$f_{\Lambda\Lambda\omega}=-9.73$   &$m_\eta=547.85$\\

$\Xi_cN\to\Sigma_c^{(*)}\Lambda$             &$...$ &$...$    &$...$                                          &$-1$                                         &$...$
      &$\frac{m_{\Xi_c}^2-m_N^2+m_\Lambda^2-m^2_{\Sigma_c^{(*)}}}{2m_{\Sigma_c^{(*)}}+2m_\Lambda}$   &$\frac{f_{\rho NN}}{g_{\rho NN}}=6.10$ &$g_{\Lambda\Lambda\eta}=-0.67$  &$m_\rho=775.49$\\

$\Xi_cN\to\Sigma_c^{(*)}\Sigma$              &$...$   &$...$     &$...$                                       &$...$                                          &$-\sqrt{2}(-\sqrt{3})$
      &$\frac{m_{\Xi_c}^2-m_N^2+m_\Sigma^2-{m_{\Sigma_c^{(*)}}}^2}{2m_{\Sigma_c^{(*)}}+2m_\Sigma}$ &$\frac{g_{\omega NN}^2}{4\pi}=20.00$ &$g_{\Lambda NK^*}=-6.08$  &$m_\omega=782.65$\\

$\Lambda_c\Sigma\to\Lambda_c\Sigma$          &$2$ &$...$  &$\sqrt{2}$   &$...$                             &$...$                                           &$0$  &$\frac{f_{\omega NN}}{g_{\omega NN}}=0.00$  &$f_{\Lambda NK^*}=-16.85$  &$m_K=495.64$\\

$\Lambda_c\Sigma\to\Xi_c^{(\prime,*)}N$      &$...$ &$...$  &$...$             &$...$                             &$1$
      &$\frac{m_{\Lambda_c}^2-m_\Sigma^2+m_N^2-m_{\Xi_c^{\prime,*}}^2}{2m_{\Xi_c^{\prime,*}}+2m_N}$  &$f_{\Lambda\Sigma\rho}=16.85$  &$g_{\Lambda\Sigma\rho}=-0.55$  &$m_{K^*}=893.61$\\

$\Lambda_c\Sigma\to\Sigma_c^{(*)}\Lambda$   &$...$ &$...$    &$...$            &$1$                             &$...$
     &$\frac{m_{\Lambda_c}^2-m_\Sigma^2+m_\Lambda^2-m_{\Sigma_c^{(*)}}^2}{2m_{\Sigma_c^{(*)}}+2m_\Lambda}$  &$g_{\Lambda\Sigma\pi}=0.67$ & &$m_{\Lambda_c}=2286.46$  \\

$\Lambda_c\Sigma\to\Sigma_c^{(*)}\Sigma$    &$...$ &$...$  &$...$    &$-\sqrt{2}$                            &$...$
    &$\frac{m_{\Lambda_c}^2-m^2_{{\Sigma_c^{(*)}}}}{2m_{\Sigma_c^{(*)}}+2m_\Sigma}$  &$g_{N\Sigma K}=0.19$ &  &$m_{\Xi_c}=2469.08$\\

$\Xi_c^{(\prime,*)}N\to\Xi_c^{(\prime,*)}N$  &$1$  &$\frac{-1}{2\sqrt{6}}$  &$\frac{1}{2\sqrt{2}}$      &$\frac{1}{2\sqrt{2}}(\frac{-3}{2\sqrt{2}})$     &$...$
    &$\frac{{m_{\Xi_c^{\prime}}^2}-m_{\Xi_c^{\prime,*}}^2}{2m_{\Xi_c^{\prime,*}}+2m_N}$ &$g_{
    \Sigma\Sigma \rho/\omega}=7.34$  &  &$m_{\Xi_c^\prime}=2578.45$\\

$\Xi_c^{(\prime,*)}N\to\Sigma_c^{(*)}\Lambda$  &$...$   &$...$  &$...$          &$...$                               &$\frac{1}{\sqrt{2}}$
    &$\frac{m_{\Xi_c^{\prime,*}}^2-m_N^2+m_\Lambda^2-m_{\Sigma_c^{(*)}}^2}{2m_{\Sigma_c^{(*)}}+2m_\Lambda}$  &$f_{\Sigma\Sigma \rho/\omega}=9.73$  &  &$m_{\Xi_c^*}=2645.63$\\

$\Xi_c^{(\prime,*)}N\to\Sigma_c^{(*)}\Sigma$   &$...$  &$...$  &$...$         &$...$                               &$1(\sqrt{\frac{3}{2}})$
    &$\frac{m_{\Xi_c^{\prime,*}}^2-m_N^2+m_\Sigma^2-m_{\Sigma_c^{(*)}}^2}{2m_{\Sigma_c^{(*)}}+2m_\Sigma}$  &$g_{
    \Sigma\Sigma \pi}=0.77$   & &$m_{\Sigma_c}=2453.46$\\

$\Sigma_c^{(*)}\Lambda\to\Sigma_c^{(*)}\Lambda$   &$1$ &$\frac{1}{\sqrt{6}}$ &$\frac{1}{\sqrt{2}}$      &$...$ &$...$                  &$\frac{m_{\Sigma_c^{(*)}}^2-m_{\Sigma_c^{(*)}}^2}{2m_{\Sigma_c^{(*)}}+2m_\Lambda}$  &$g_{
    \Sigma\Sigma \eta}=0.67$   & &$m_{\Sigma_c^*}=2518.10$\\

$\Sigma_c^{(*)}\Lambda\to\Xi_c^*N$   &$...$ &$...$  &$...$  &$...$  &$\frac{1}{\sqrt{2}}$      &$\frac{m_{\Sigma_c^{(*)}}^2-m_\Lambda^2+m_N^2-m_{\Xi_c^{*}}^2}{2m_{\Xi_c^{*}}+2m_N}$  &$g_{
    N\Sigma K^*}=-4.15$   & &$m_{\Lambda}=1115.68$ \\

$\Sigma_c^{(*)}\Lambda\to\Sigma_c^{(*)}\Sigma$
                                    &$...$ &$...$  &$...$  &$1$   &$...$ &$\frac{m_{\Sigma_c^{(*)}}^2-m_\Lambda^2+m_\Sigma^2-m_{\Sigma_c^{(*)}}^2}{2m_{\Sigma_c^{(*)}}+2m_\Sigma}$ &$f_{
    N\Sigma K^*}=9.73$   & &$m_{\Sigma}=1193.15$\\

$\Sigma_c^{(*)}\Sigma\to\Sigma_c^{(*)}\Sigma$
                                       &$1$  &$\frac{1}{\sqrt{6}}$ &$\frac{1}{\sqrt{2}}$   &$\frac{-1}{\sqrt{2}}(-\sqrt{2})$   &$...$   &$\frac{m_{\Sigma_c^{(*)}}^2-m_{\Sigma_c^{(*)}}^2}{2m_{\Sigma_c^{(*)}}+2m_\Sigma}$  &$g_{
    \Lambda N K}=-1.00$   & &$m_{N}=938.919$\\
\bottomrule[1pt]
\bottomrule[1pt]
\end{tabular}
\end{table*}

\renewcommand\tabcolsep{0.3cm}
\renewcommand{\arraystretch}{1.7}
\begin{table}[!htbb]
\caption{Matrix elements for the spin-spin interactions and tensor force operators in the OBE effective potentials. Here, $\langle \bm{\sigma_1}\cdot\bm{\epsilon_4^{n\dag}}\rangle=-\langle \bm{\sigma_1}\cdot(i\bm{\sigma_2}\times\bm{\epsilon_4^{n\dag}})\rangle$, $\langle\bm{\sigma}_2\cdot\bm{\epsilon}_4^{n\dag}\rangle=0$.}\label{operator}
{\begin{tabular}{cll}
\toprule[1pt]
\toprule[1pt]
{Spin}
       &$\langle\bm{\sigma_1}\cdot\bm{\sigma_2}\rangle$       &$\langle S(\hat{r},\bm{\sigma_1},\bm{\sigma_2})\rangle$\\\hline
$J=0$   &$\left(-3\right)$
              &$\left(0\right)$\\
$J=1$    &$\left(\begin{array}{cc}1  &0\\  0   &1 \end{array}\right)$
             &{$\left(\begin{array}{cc}0  &\sqrt{8}\\  \sqrt{8}   &-2 \end{array}\right)$}
                  \\
\bottomrule[1pt]
{Spin}
       &$\langle \bm{\epsilon_2^n}\cdot\bm{\epsilon_4^{b\dag}}\rangle$
       &$\langle i\bm{\sigma_1}\cdot\left(\bm{\epsilon_2^n}\times\bm{\epsilon_4^{b\dag}}\right)\rangle$       \\\hline
$J=0$   &$\left(1\right)$
                             &$\left(1\right)$
                                 \\
$J=1$    &$\left(\begin{array}{ccc} 1 & 0 & 0 \\ 0 & 1 & 0 \\ 0 & 0 & 1\end{array}\right)$
                          &{$\left(\begin{array}{ccc} -\frac{5}{3} & 0 & 0 \\ 0 & -\frac{5}{3} & 0 \\ 0 & 0 & 1\end{array}\right)$}
                                 \\
$J=2$    &$\left(\begin{array}{ccc} 1 & 0 & 0 \\ 0 & 1 & 0 \\ 0 & 0 & 1\end{array}\right)$
               &$\left(\begin{array}{ccc} 1 & 0 & 0 \\ 0 & -\frac{5}{3} & 0 \\ 0 & 0 & 1\end{array}\right)$\\
\bottomrule[1pt]
{Spin}
       &$\langle S(\hat{r},\bm{\sigma_1},i\bm{\epsilon_2^n}\times\bm{\epsilon_4^{b\dag}})\rangle$             &$\langle \bm{\sigma_1}\cdot\bm{\epsilon_4^{n\dag}}\rangle$
       \\\hline
$J=0$   &$\left(-2\right)$
               &$\left(0\right)$
           \\
$J=1$    &{$\left(\begin{array}{ccc} 0 & -\frac{\sqrt{2}}{3} & -\sqrt{2} \\ -\frac{\sqrt{2}}{3} & \frac{1}{3} & -1 \\-\sqrt{2} & -1 & -1\end{array}\right)$}
                                    &$\left(\begin{array}{cc} 2 \sqrt{\frac{2}{3}} & 0 \\ 0 & 2 \sqrt{\frac{2}{3}} \\ 0 & 0 \\\end{array}\right)$
                        \\
$J=2$      &{$\left(\begin{array}{ccc} 0 & \sqrt{\frac{6}{5}} & \sqrt{\frac{14}{5}} \\ \sqrt{\frac{6}{5}} & -\frac{1}{3} & -\sqrt{\frac{3}{7}} \\ \sqrt{\frac{14}{5}} & -\sqrt{\frac{3}{7}} & \frac{3}{7}\end{array}\right)$}
                       &
\\\bottomrule[1pt]
{Spin}
       &$\langle S(\hat{r},\bm{\sigma_1},\bm{\epsilon_4^{n\dag}})\rangle$
       &$\langle S(\hat{r},\bm{\sigma_1},i\bm{\sigma_2}\times\bm{\epsilon_4^{n\dag}})\rangle$\\\hline
$J=0$   &$\left(\sqrt{6}\right)$
              &$\left(0\right)$\\
$J=1$    &$\left(\begin{array}{cc} 0 & -\frac{1}{\sqrt{3}} \\ -\frac{1}{\sqrt{3}} & \frac{1}{\sqrt{6}} \\ -\sqrt{3} & -\sqrt{\frac{3}{2}} \\\end{array}\right)$
                             &$\left(\begin{array}{cc} 0 & \frac{1}{\sqrt{3}} \\ \frac{1}{\sqrt{3}} & -\frac{1}{\sqrt{6}} \\ \sqrt{3} & \sqrt{\frac{3}{2}} \\\end{array}\right)$
                             \\\bottomrule[1pt]\bottomrule[1pt]
\end{tabular}}
\end{table}

\section{Numerical results}\label{sec3}

Using the deduced OBE effective potentials, we numerically explore the properties of the $\Xi_c^{(',*)}N$ systems and search for possible bound state solutions by solving the coupled-channel Schr\"{o}dinger equations. When considering loosely bound molecular states, effective solutions generally exhibit certain key characteristics. For a single-channel scenario, the size of the $S$-wave molecular state $R$ can be approximated using the binding energy $E$ and the reduced mass $\mu$, with $R\sim1/\sqrt{2\mu E}$. For a loosely bound molecular state, its size should exceed that of all constituent components, implying that the binding energy falls within the range of several MeV to several tens of MeV. In a coupled-channel scenario, the size of the coupled molecule can be estimated using the reduced mass for the dominant channel $\mu_D$ and the energy $\tilde{E}=M_{L}-M_D+E$ measured from the dominant channel, with $R\sim1/\sqrt{2\mu_D \tilde E}$. Here, $M_L$ and $M_D$ represent the masses of the lowest and dominant channels, respectively. If the dominant channel is not the lowest channel, $R$ may be smaller than the size of all components, making it an unsuitable candidate for a good molecular state. In this study, we will focus on searching for loosely bound states primarily composed of the lowest channels. Furthermore, based on experience with nucleon-nucleon interactions, a reasonable cutoff value is approximately 1.00 GeV. Consequently, a bound state that meets these criteria, along with the reasonable cutoff value, can be considered a promising hadronic molecular candidate.

\subsection{The $\Xi_cN$ systems}

We first search for potential charm-strange deuteronlike hexaquarks mainly composed of the $\Xi_cN$ system. As is former emphasized, in the single-channel case, the OPE interactions are absent because the $\Xi_c\Xi_c\pi$ coupling is suppressed by the spin-parity conservation for the light quarks in the heavy quark symmetry. The $\sigma$ exchange interaction contributes an attractive force. The $\omega$ exchange exerts a repulsive force. The $\rho$ exchange interaction is attractive in the isoscalar $\Xi_cN$ system, while it is 3 times weaker repulsive in the isovector $\Xi_cN$ system. When varying the cutoff in the range of $\Lambda$ smaller than 2.00 GeV, we cannot find bound state solutions for the single $\Xi_cN$ systems with $I(J^P)=0(0^+)$, $0(1^+)$, $1(0^+)$, and $1(1^+)$. Therefore, the OBE effective potentials cannot provide strong attractive interactions to bind the single $\Xi_cN$ systems.

After that, we further adopt the coupled-channel effects while considering the coupling from $\Xi_c^{(',*)}N$, $\Lambda_c\Sigma$, $\Sigma_c^{(*)}\Lambda$, and $\Sigma_c^{(*)}\Sigma$ channels. Furthermore, the OBE interactions become much more complicated; especially the long-range interactions from the OPE process can exist in the off-diagonal elements of the coupled-channel effective potentials. In our coupled-channel analysis, we only present the numerical results with only considering the $S$-wave contributions, because we find the contributions from the $D$-wave play an minor role in the formative of the bound states. Finally, the discussed channels for the coupled $\Xi_c^{(',*)}N/\Lambda_c\Sigma/\Sigma_c^{(*)}\Lambda/\Sigma_c^{(*)}\Sigma$ channel systems include
\begin{eqnarray}\left.\begin{array}{ll}
0(0^+): &\Xi_cN, \Xi_c^{\prime}N, \Sigma_c\Sigma,\\
1(0^+): &\Xi_cN, \Lambda_c\Sigma, \Xi_c^{\prime}N, \Sigma_c\Lambda, \Sigma_c\Sigma,\\
0(1^+): &\Xi_cN, \Xi_c^{\prime}N, \Xi_c^*N, \Sigma_c\Sigma, \Sigma_c^*\Sigma,\\
1(1^+): &\Xi_cN, \Lambda_c\Sigma, \Xi_c^{\prime}N, \Sigma_c\Lambda, \Xi_c^*N, \Sigma_c^*\Lambda, \Sigma_c\Sigma, \Sigma_c^*\Sigma.\end{array}\right.
\end{eqnarray}

In Table \ref{num1}, we collect the bound state solutions for the coupled $\Xi_cN/\Lambda_c\Sigma/\Xi_c^{\prime}N/\Sigma_c\Lambda/\Xi_c^*N/\Sigma_c^*\Lambda/\Sigma_c\Sigma/\Sigma_c^*\Sigma$ channel systems with $I(J^P)=0(0^+)$, $1(0^+)$, $0(1^+)$, and $1(1^+)$, including the binding energy $E$, the rms radius $r_{rms}$, and the probabilities for the discussed channels. When we take the cutoff in the range of $\Lambda<2.00$ GeV, we can obtain four bound states. Obviously, the bound state properties for the coupled state with $0(1^+)$ satisfy the features for a reasonable loosely bound molecular state, and the $\Xi_cN$ components are the dominant channels. For the remaining three states, they cannot be recommended as good $\Xi_cN$ molecules as the dominant channels are not the $\Xi_cN$. However, as their OBE interactions can be strong enough attractive interactions, it is possible to search for charm-strange hexaquarks resonances near the mass thresholds for the corresponding dominant channels, which we shall leave to future study.

\begin{table*}[!hbtp]
\renewcommand\tabcolsep{0.7cm}
\renewcommand{\arraystretch}{1.6}
\caption{The bound state solutions for the coupled $\Xi_cN, \Lambda_c\Sigma/\Xi_c^{\prime}N/\Sigma_c\Lambda/\Xi_c^*N/\Sigma_c^*\Lambda/\Sigma_c\Sigma/\Sigma_c^*\Sigma$ channel systems with $I(J^P)=0(0^+)$, $1(0^+)$, $0(1^+)$, and $1(1^+)$. The cutoff $\Lambda$, the binding energy $E$, and the root-mean-square $r_{\text{rms}}$ are in units of GeV, MeV, and fm, respectively.} \label{num1}
\begin{tabular}{lrrrrrr}
\toprule[1pt]
\toprule[1pt]
  &\multicolumn{3}{c}{$I(J^P)=0(0^{+})$}                                    & \multicolumn{3}{c}{$I(J^P)=1(0^{+})$} \\
  \cline{1-7}
  $\Lambda$                         &$0.78$       &$0.79$     &$0.80$         &$0.79$      &$0.80$     &$0.81$    \\
  $E$                               &$-44.63$     &$-54.72$   &$-66.05$       &$-2.98$     &$-11.65$   &$-22.86$  \\
  $r$                               &$0.71$       &$0.67$     &$0.63$         &$1.57$      &$0.76$     &$0.61$     \\
  $\Xi_cN$               &$48.96$      &$46.33$    &$43.96$        &$39.23$     &$17.68$    &$10.80$    \\
  $\Lambda_c^{}\Sigma$   &...          &...        &...            &$24.15$     &$33.00$    &$35.28$     \\
  $\Xi_c^{\prime}N$      &$51.00$      &$53.65$    &$56.04$        &$2.80$      &$3.00$     &$2.75$     \\
  $\Sigma_c\Lambda$      &...          &...        &...            &$16.10$     &$21.28$    &$22.90$     \\
  $\Sigma_c^{}\Sigma$    &$0.05$       &$0.02$     &$\sim0$        &$17.72$     &$25.04$    &$28.27$      \\
   &\multicolumn{3}{c}{$I(J^P)=0(1^{+})$}                                    & \multicolumn{3}{c}{$I(J^P)=1(1^{+})$} \\
 \cline{2-4} \cline{5-7}
  $\Lambda$                           &$0.78$      &$0.79$      &$0.80$                      &$0.80$    &$0.81$    &$0.82$     \\
  $E$                                 &$-27.99$    &$-36.69$    &$-46.75$                    &$-0.16$   &$-7.96$   &$-19.91$              \\
  $r$                                 &$0.86$      &$0.77$      &$0.71$                      &$3.32$    &$0.90$    &$0.63$       \\
  $\Xi_cN$                 &$63.91$     &$60.10$     &$56.57$                     &$70.38$   &$21.72$   &$11.34$   \\
  $\Lambda_c\Sigma$        &...         &...         &...                         &$14.12$   &$37.84$   &$41.85$ \\
  $\Xi_c^{\prime}N$        &$5.38$      &$5.89$      &$6.33$                      &$0.17$    &$0.30$    &$0.28$ \\
  $\Sigma_c\Lambda$        &...         &...         &...                     &$0.94$    &$2.38$    &$2.68$ \\
  $\Xi_c^{*}N$             &$30.64$     &$33.97$     &$37.09$                     &$0.95$    &$1.77$    &$1.71$ \\
  $\Sigma_c^*\Lambda$      &...         &...         &...                         &$5.67$    &$14.52$   &$16.59$ \\
  $\Sigma_c\Sigma$         &$0.01$      &$0.01$      &$\sim0$                     &$1.03$    &$2.85$    &$3.37$ \\
  $\Sigma_c^*\Sigma$       &$0.06$      &$0.03$      &$0.01$                      &$6.74$    &$18.63$   &$22.19$ \\
\midrule[1pt]
\midrule[1pt]
\end{tabular}
\end{table*}

From the current results, we can predict a charm-strange molecular candidate, which is mainly made up by the $\Xi_cN$ component with $I(I^P)=0(1^+)$. The coupled-channel effects play a crucial role.

\subsection{The $\Xi_c^{\prime}N$ systems}\label{result2}

\begin{table*}[!hbtp]
\renewcommand\tabcolsep{0.45cm}
\renewcommand{\arraystretch}{1.7}
\caption{The bound state solutions for the single $\Xi_c^{\prime}N$ systems and the coupled $\Xi_c^{\prime}N/\Sigma_c\Lambda/\Xi_c^*N/\Sigma_c^*\Lambda/\Sigma_c\Sigma/\Sigma_c^*\Sigma$ channel systems with $I(J^P)=0(0^+)$, $1(0^+)$, $0(1^+)$, and $1(1^+)$. The cutoff $\Lambda$, the binding energy $E$, and the root-mean-square $r_{\text{rms}}$ are in units of GeV, MeV, and fm, respectively.} \label{num2}
\begin{tabular}{cccccccc}
\toprule[1pt]
\toprule[1pt]
 Single channel &&\multicolumn{3}{c}{$I(J^P)=0(0^{+})$}                                      & \multicolumn{3}{c}{$I(J^P)=1(0^{+})$} \\
\cline{3-8}
&$\Lambda$                       &$0.78$        &$0.79$      &$0.80$     &\ldots&\ldots&\ldots \\
&$E$                             &$-37.44$      &$-43.16$    &$-49.41$          &\ldots&\ldots&\ldots                                \\
&$r$                             &$0.90$        &$0.85$      &$0.81$            &\ldots&\ldots&\ldots                                  \\
&$\Xi_c^{\prime}N({}^1S_{0})$    &$100.00$      &$100.00$    &$100.00$          &\ldots&\ldots&\ldots                                     \\
  &&\multicolumn{3}{c}{$I(J^P)=0(1^{+})$}                                      & \multicolumn{3}{c}{$I(J^P)=1(1^{+})$} \\
\cline{3-8}
&$\Lambda$                        &$1.30$      &$1.60$      &$1.9$       &$1.08$  &$1.14$         &$1.20$     \\
&$E$                              &$-0.65$      &$-2.26$    &$-5.36$      &$-0.86$  &$-3.25$        &$-10.23$              \\
&$r$                               &$3.86$      &$3.03$     &$2.28$       &$3.37$    &$2.28$      &$1.37$       \\
&$\Xi_c^{\prime}N({}^3S_{1})$    &$97.45$      &$96.98$     &$96.87$     &$99.78$  &$99.84$   &$99.90$ \\
&$\Xi_c^{\prime}N({}^3D_{1})$    &$2.55$      &$3.02$      &$3.13$       &$0.22$  &$0.16$   &$0.10$ \\
\midrule[1pt]
Coupled channel  & &\multicolumn{3}{c}{$I(J^P)=0(0^{+})$}                                    & \multicolumn{3}{c}{$I(J^P)=1(0^{+})$} \\
  \cline{3-8}
 &$\Lambda$                              &$0.78$       &$0.79$      &$0.80$          &$0.84$   &$0.85$   &$0.86$      \\
&$E$                                    &$-38.00$     &$-43.93$    &$-50.44$        &$-5.08$  &$-13.20$  &$-22.43$    \\
&$r$                                    &$0.89$       &$0.85$      &$0.80$          &$0.96$    &$0.70$ &$0.63$        \\
&$\Xi_c^{\prime}N$           &$99.78$      &$99.61$     &$99.38$         &$15.17$    &$8.51$   &$6.16$ \\
&$\Sigma_c\Lambda$           &...          &...         &...             &$46.73$   &$48.79$ &$48.60$\\
&$\Sigma_c^{}\Sigma$         &$0.22$       &$0.39$      &$0.62$          &$38.10$  &$42.70$   &$45.23$ \\
  & &\multicolumn{3}{c}{$I(J^P)=0(1^{+})$}                                    & \multicolumn{3}{c}{$I(J^P)=1(1^{+})$} \\
 \cline{3-8}
 &$\Lambda$                            &$0.80$      &$0.81$    &$0.82$             &$0.89$      &$0.90$      &$0.91$     \\
&$E$                                  &$-1.66$     &$-6.31$   &$-12.68$           &$-2.14$     &$-14.55$    &$-28.35$              \\
&$r$                                  &$2.45$      &$1.33$    &$0.97$             &$1.12$      &$0.54$      &$0.51$     \\
&$\Xi_c^{\prime}N$         &$64.94$     &$45.28$   &$35.03$            &$13.65$     &$1.73$      &$0.93$ \\
&$\Sigma_c\Lambda$         &...         &...       &...                &$8.24$      &$8.89$      &$8.23$ \\
&$\Xi_c^{*}N$              &$34.84$    &$54.17$     &$64.00$           & $2.03$     &$2.58$      &$2.58$ \\
&$\Sigma_c^*\Lambda$       &...         &...       &...                &$32.95$    &$36.88$     &$37.04$ \\
&$\Sigma_c\Sigma$          &$0.02$      &$0.06$    &$0.13$             &$6.45$     &$7.33$      &$7.43$ \\
&$\Sigma_c^*\Sigma$        &$0.20$      &$0.49$    &$0.84$             &$36.68$    &$42.59$     &$43.79$ \\
\bottomrule[1pt]
\bottomrule[1pt]
\end{tabular}
\end{table*}

\begin{table*}[!hbtp]
\renewcommand\tabcolsep{0.45cm}
\renewcommand{\arraystretch}{1.7}
\caption{The bound state solutions for the single $\Xi_c^*N$ systems and the coupled $\Xi_c^*N/\Sigma_c^*\Lambda/\Sigma_c\Sigma/\Sigma_c^*\Sigma$ channel systems with $I(J^P)=0(1^+)$, $1(1^+)$, $0(2^+)$, and $1(2^+)$. The cutoff $\Lambda$, the binding energy $E$, and the root-mean-square $r_{\text{rms}}$ are in units of GeV, MeV, and fm, respectively.} \label{num3}
\begin{tabular}{cccccccc}
\toprule[1pt]
\toprule[1pt]
Single channel &&\multicolumn{3}{c}{$I(J^P)=0(1^{+})$}                                      & \multicolumn{3}{c}{$I(J^P)=1(1^{+})$} \\
\cline{3-8}
&$\Lambda$                 &$0.78$      &$0.79$     &$0.80$           &\ldots&\ldots &\ldots                      \\
&$E$                       &$-25.31$    &$-29.64$   &$-34.31$         &\ldots&\ldots &\ldots      \\
&$r$                       &$1.07$      &$1.00$     &$0.96$           &\ldots&\ldots &\ldots      \\
&$\Xi_c^{*}N({}^3S_{1})$   &$98.61$     &$98.63$    &$98.65$          &\ldots&\ldots &\ldots\\
&$\Xi_c^{*}N({}^3D_{1})$   &$0.13$      &$0.13$     &$0.13$           &\ldots&\ldots &\ldots\\
&$\Xi_c^{*}N({}^5D_{1})$   &$1.26$      &$1.24$     &$1.22$           &\ldots&\ldots &\ldots\\
\cline{2-8}
&&\multicolumn{3}{c}{$I(J^P)=0(2^{+})$}                                      & \multicolumn{3}{c}{$I(J^P)=1(2^{+})$} \\
\cline{3-8}
&$\Lambda$                     &$1.10$      &$1.50$      &$1.90$         &$0.94$    &$1.02$    &$1.10$     \\
&$E$                           &$-0.39$     &$-3.63$     &$-11.16$       &$-0.46$   &$-2.89$   &$-10.00$           \\
&$r$                           &$4.08$      &$2.65$      &$1.78$         &$3.70$    &$2.43$    &$1.42$       \\
&$\Xi_c^{*}N({}^5S_{2})$       &$96.72$     &$95.33$     &$95.64$       &$99.63$    &$99.66$   &$99.84$ \\
&$\Xi_c^{*}N({}^3D_{2})$       &$0.71$      &$0.86$      &$0.71$        &$0.12$     &$0.11$    &$0.05$ \\
&$\Xi_c^{*}N({}^5D_{2})$       &$2.57$      &$3.80$      &$3.65$        &$0.25$     &$0.23$    &$0.11$ \\
\bottomrule[1pt]
Coupled channel &&\multicolumn{3}{c}{$I(J^P)=0(1^{+})$}                                    & \multicolumn{3}{c}{$I(J^P)=1(1^{+})$} \\
  \cline{3-8}
&$\Lambda$                        &$0.78$      &$0.79$      &$0.80$         &$0.83$     &$0.84$    &$0.85$     \\
&$E$                              &$-20.68$    &$-24.83$   &$-29.43$        &$-0.02$    &$-6.11$   &$-13.78$              \\
&$r$                              &$1.11$      &$1.04$     &$0.97$          &$2.80$     &$0.87$    &$0.70$     \\
&$\Xi_c^{*}N$          &$99.77$     &$99.56$    &$99.21$         &$44.70$    &$11.48$   &$6.84$\\
&$\Sigma_c^*\Lambda$   &...         &...        &...             &$29.11$    &$45.08$   &$46.10$ \\
&$\Sigma_c\Sigma$      &$0.14$      &$0.19$     &$0.27$          &$8.10$     &$13.00$   &$13.63$ \\
&$\Sigma_c^*\Sigma$    &$0.09$      &$0.25$     &$0.53$          &$18.09$    &$30.44$   &$33.44$ \\
  \cline{2-8}
   &&\multicolumn{3}{c}{$I(J^P)=0(2^{+})$}                                    & \multicolumn{3}{c}{$I(J^P)=1(2^{+})$} \\
 \cline{3-8}
&$\Lambda$                        &$1.50$      &$1.65$     &$1.80$           &$0.95$     &$1.00$   &$1.05$     \\
&$E$                              &$-0.61$     &$-4.62$    &$-13.53$         &$-0.94$    &$-3.89$  &$-11.51$              \\
&$r$                              &$3.83$      &$2.27$     &$1.46$           &$3.32$     &$2.13$   &$1.29$     \\
&$\Xi_c^{*}N$          &$98.96$     &$94.64$    &$83.86$          &$99.31$    &$97.69$  &$91.83$\\
&$\Sigma_c^*\Lambda$   &...         &...        &...              &$0.65$     &$2.16$   &$7.23$ \\
&$\Sigma_c^*\Sigma$    &$1.04$      &$5.36$      &$16.14$         &$0.03$     &$0.15$   &$0.94$ \\
\bottomrule[1pt]
\bottomrule[1pt]
\end{tabular}
\end{table*}

Following in the same procedures, we next investigate the possible existence of the charm-strange hexaquark molecules dominated by the $\Xi_c^{\prime}N$ channels. In Table \ref{num2}, we present the bound state solutions for the single $\Xi_c^{\prime}N$ systems and the coupled $\Xi_c^{\prime}N/\Sigma_c\Lambda/\Xi_c^*N/\Sigma_c^*\Lambda/\Sigma_c\Sigma/\Sigma_c^*\Sigma$ channel systems with $I(J^P)=0(0^+)$, $1(0^+)$, $0(1^+)$, and $1(1^+)$. Compared to the single $\Xi_cN$ systems, there exist the long-range interactions from the OPE interactions. When we only consider the $S$-$D$-wave mixing effects, one cannot find the bound state solutions for the single $\Xi_c^{\prime}N$ system with $I(J^P)=1(0^+)$ in the cutoff range of $\Lambda<2.00$ GeV. For the remaining three systems, the loosely bound state solutions for the $\Xi_c^{\prime}N$ states with the $0(0^+)$, $0(1^+)$, and $1(1^+)$ emerge as the cutoffs are taken around 1.00 GeV. And the $S$-wave $\Xi_c^{\prime}N$ components are the dominant channels. In summary, since the cutoff values for these bound states are close to those found in nucleon-nucleon interactions \cite{Tornqvist:1993ng,Tornqvist:1993vu}, and their rms radii $r_{rms}$ are around or larger than 1.00 fm, the single $\Xi_c^{\prime}N$ states with $I(I^P)=0(0^+)$, $0(1^+)$, and $1(1^+)$ can be considered as good hadronic molecular candidates.

After delving into the coupled-channel effects, there can exist four bound states with the cutoff taking around 1.00 GeV, i.e., the coupled $\Xi_c^{\prime}N/\Sigma_c\Lambda/\Xi_c^*N/\Sigma_c^*\Lambda/\Sigma_c\Sigma/\Sigma_c^*\Sigma$ channel states with $I(J^P)=0(0^+)$, $1(0^+)$, $0(1^+)$, and $1(1^+)$. Furthermore, the bound state properties for the coupled $\Xi_c^{\prime}N/\Sigma_c\Lambda/\Xi_c^*N/\Sigma_c^*\Lambda/\Sigma_c\Sigma/\Sigma_c^*\Sigma$ channel states with $I(J^P)=0(0^+)$ do not change too much in comparison with those in the single $\Xi_c^{\prime}N$ bound state with $0(0^+)$. Nevertheless, the coupled-channel effects positively contribute to the binding of this coupled bound state due to the smaller cutoff value, and the $\Xi_c^{'}N({}^{3}S_{1})$ are still the dominant channels.

For the coupled $\Xi_c^{\prime}N/\Sigma_c\Lambda/\Xi_c^*N/\Sigma_c^*\Lambda/\Sigma_c\Sigma/\Sigma_c^*\Sigma$ channel state with $I(J^P)=0(1^+)$, it binds deeper than the single $\Xi_c^{\prime}N$ state with $0(1^+)$. With the increasing of the binding energy, the $\Xi_c^*N$ channel becomes more important than the $\Xi_c^{\prime}N$ channel. This indicates that the $\Xi_c^*N$ interactions may be strongly attractive enough to form possible hadronic molecular candidate, which we discuss in the next subsection.

For the coupled $\Xi_c^{\prime}N/\Sigma_c\Lambda/\Xi_c^*N/\Sigma_c^*\Lambda/\Sigma_c\Sigma/\Sigma_c^*\Sigma$ channel states with $I(J^P)=1(0^+)$ and $1(1^+)$, the probabilities for the $\Xi_c^{\prime}N$ are small, but the other channels composed of the charmed baryon and strange baryon are very important. It is possible to search for hexaquark resonances with $1(0^+)$ and $1(1^+)$ near the $\Sigma_c^{(*)}\Lambda(\Sigma)$ thresholds. In Ref. \cite{Kong:2022rvd}, authors predicted the existence of these states by solving the quasipotential Bethe-Salpeter equations.

\subsection{The $\Xi_c^*N$ systems}\label{result3}

For the $\Xi_c^*N$ systems, the quantum number configurations discussed are $I(J^P)=0(1^+)$, $0(2^+)$, $1(1^+)$, and $1(2^+)$. In Table \ref{num3}, we present the bound state solutions for the single $\Xi_c^*N$ systems and the coupled $\Xi_c^*N/\Sigma_c^*\Lambda/\Sigma_c\Sigma/\Sigma_c^*\Sigma$ channel systems with $I(J^P)=0(1^+)$, $1(1^+)$, $0(2^+)$, and $1(2^+)$. We can obtain the weakly bound energies and reasonable rms radii for the single $\Xi_c^*N$ systems with $0(1^+)$, $0(2^+)$, and $0(2^+)$ when the cutoff values fall into the reasonable cutoff region. Thus, they can be suggested as good molecular candidates; whereas for the single $\Xi_c^*N$ systems with $1(1^+)$, the OBE interactions are too weak to form the loosely bound states.

When we consider the coupled-channel effects, as shown in Table \ref{num3}, we find that the $\Xi_c^*N$ states with $0(1^+)$ and $1(2^+)$ bind deeper, and the dominant channels are still the $\Xi_c^*N$. Whereas for the coupled $\Xi_c^*N/\Sigma_c^*\Lambda/\Sigma_c\Sigma/\Sigma_c^*\Sigma$ channel state with $0(2^+)$, the OBE interactions become a little weaker, nevertheless, it can be still regarded as a good molecular candidate as we can still obtain the loosely bound state properties at the reasonable cutoff value. In addition, the bound state solutions for the coupled $\Xi_c^*N/\Sigma_c^*\Lambda/\Sigma_c\Sigma/\Sigma_c^*\Sigma$ channel with $1(1^+)$ emerge with the reasonable cutoff $\Lambda$; however, the dominant channels are the $\Sigma_c^*\Lambda$ and $\Sigma_c^*\Sigma$ as it binds deeper and deeper.

All in all, we can predict the single $\Xi_c^*N$ states with $0(1^+)$, $0(2^+)$, and $1(2^+)$ as good charm-strange hexaquark molecular candidates.

\section{Summary}\label{sec4}

The study of the hadron-hadron interactions can provide invaluable insights into the intrinsic structures of the new hadron states. With the continuous increase in experimental energy levels and the accumulation of experimental data, one can expect further breakthroughs in the experimental study of these new hadronic states. In this work, we systematically study the interactions between charm-strange baryons in the $S$-wave configuration and nucleons using the OBE model and considering the $S$-$D$-wave mixing effects. In addition, we further take into account the and the coupled-channel effects from the charmed baryon and the hyperon interactions.

Finally, we find the coupled $\Xi_cN/\Xi_c^{\prime}N/\Xi_c^*N/\Sigma_c\Sigma/\Sigma_c^*\Sigma$ channel state with $I(J^P)=0(1^+)$ can be recommended as the prime molecular candidate, the dominant channel is the $\Xi_cN$, followed by the $\Xi_c^*N$ channel, and the coupled-channel effects play an important role. In addition, there can exist several single-channel charm-strange hexaquarks molecules, the $\Xi_c^{\prime}N$ states with $I(I^P)=0(0^+)$, $0(1^+)$, and $1(1^+)$ and the $\Xi_c^*N$ states with $0(1^+)$, $0(2^+)$, and $1(2^+)$.
We hope the predicted mass spectrum behaviors can provide valuable information for the search for the charmed hypernuclei in future experiments.

\section*{ACKNOWLEDGMENTS}

R.C. is supported by the National Natural Science Foundation of China under Grant No. 12305139 and the Xiaoxiang Scholars Programme of Hunan Normal University. X.L. is supported by the National Natural Science Foundation of China under Grants No. 12335001 and No. 12247101, National Key Research and Development Program of China under Contract No. 2020YFA0406400, the 111 Project under Grant No. B20063, the fundamental Research Funds for the Central Universities, and the project for top-notch innovative talents of Gansu province.

\end{document}